\documentstyle[epsf]{elsart}
\begin{document}
\journal{Physica A}
\begin{frontmatter}
\title{
Simulation of a directed random-walk model: the effect of
pseudo-random-number correlations}
\author[Landau]{L.N. Shchur},
\author[Delft]{J.R. Heringa} and
\author[Delft]{H.W.J. Bl\"ote}
\address[Landau]{
Landau Institute for Theoretical Physics, 117940 GSP-1 Moscow V-334,
Russia}
\address[Delft]{
Department of Applied Physics, Delft University of Technology,\\
Lorentzweg 1, 2628 CJ Delft, The Netherlands}
\begin{abstract}

\noindent
We investigate the mechanism that leads to systematic deviations
in cluster Monte Carlo simulations when correlated pseudo-random
numbers are used.
We present a simple model, which enables an analysis of the
effects due to correlations in several types
of pseudo-random-number sequences.
This model provides qualitative understanding of the bias
mechanism in a class of cluster Monte Carlo algorithms.
\end{abstract}
\keyword
random-number generator, directed random walk
\endkeyword
\end{frontmatter}
\newpage
\section{Introduction}
\label{sec_intro}

The Monte Carlo method to obtain statistical averages for a model is
widely used when exact calculations are not available.
The error, expected from a standard statistical analysis,
is proportional to the inverse square root of the number
of randomly chosen samples.
One may find biased results if the samples are not representative.
Upper bounds to the error are not known in general.
Standard statistical analysis of simulation data suffices to
obtain the errors of a random statistical nature.
Statistically independent random numbers do not introduce a bias in
Monte Carlo simulations.
However, it is not possible to achieve independence
with a deterministic recipe, as is commonly used in Monte Carlo
calculations\cite{HN}.
Although one generates seemingly irregular numbers
the underlying production rule induces relations or correlations between
the pseudo-random numbers.
The production rule itself
already constitutes a correlation between the generated
pseudo-random numbers.
The correlations may
cause a bias in the simulation results that is difficult to assess.

Substitution of the production rule in itself leads to further
correlations involving more random numbers.
By repeated substitutions one finds a hierarchy of correlations.
Such hierarchies were studied by Compagner\cite{AC1,AC2}.
Due to the multitude of possible correlations it is not feasible to study
all of them.
One has to analyse
which correlations are detrimental to the intended use
of the pseudo-random numbers.
Compagner notes, that correlations between few numbers or
closely spaced numbers are most important.
The effect of such correlations has actually been observed
in Metropolis type Monte Carlo simulations of Ising
models \cite{DISP,Wilding}.

In this work, we investigate the consequences of correlations between
pseudorandom numbers in an example of a simple random-walk model.
An analysis of the effect of the correlations present
when a given generator is used may guide us in the choice
of an appropriate production rule.
In Section \ref{sec_model} we introduce our stochastic walk model.
In Section \ref{sec_rng} we discuss the class of generators we used.
Deviations caused by the correlations are described in
Section \ref{sec_core} for shift register generators and in
Section \ref{sec_fib} for the lagged Fibonacci random number generator.
In the last Section we give a discussion and conclusion.

\section{The walk model}
\label{sec_model}

The Ising simulations using cluster updates with the Wolff \cite{Wolff}
method seem to be very sensitive to deficiencies in a number of
pseudo-random-number generators \cite{FLW,SB}.
The crucial element of such simulations is the formation of clusters.
These are formed by joining spins which are connected
by `active' bonds.
In ferromagnetic models bonds can only be active,
if the spins on either side of the bond are in the same state.
The probability, that a bond is active,
still depends on the coupling between the spins.

In order to study the random-number generator induced bias
we replace the Wolff cluster formation process by a simpler one.
First we place the Ising spins on a one-dimensional lattice.
Second the cluster is grown only in one direction
(or, equivalently, the cluster formation always begins at an open
boundary of the spin chain).
Third the cluster is grown in a
configuration of parallel spins only.
This third simplification is less far reaching than it may seem.
It does not modify the Ising statistics of the system.
The probability that Ising spins, coupled with a strength $K$,
are parallel is $\left(1+e^{-2K}\right)^{-1}$.
If they are parallel the random-cluster model yields a probability
$(1-e^{-2K})$ that they are connected by an active bond,
i.e. that the two spins belong to the same cluster.
Thus, in Wolff simulations of the one-dimensional Ising model
the probability that each next spin is included in the cluster,
equals $\mu=\tanh(K)$: a constant depending on $K$ only, just
as in our simplified model.

\begin{figure}
\epsfxsize=\textwidth
\epsffile{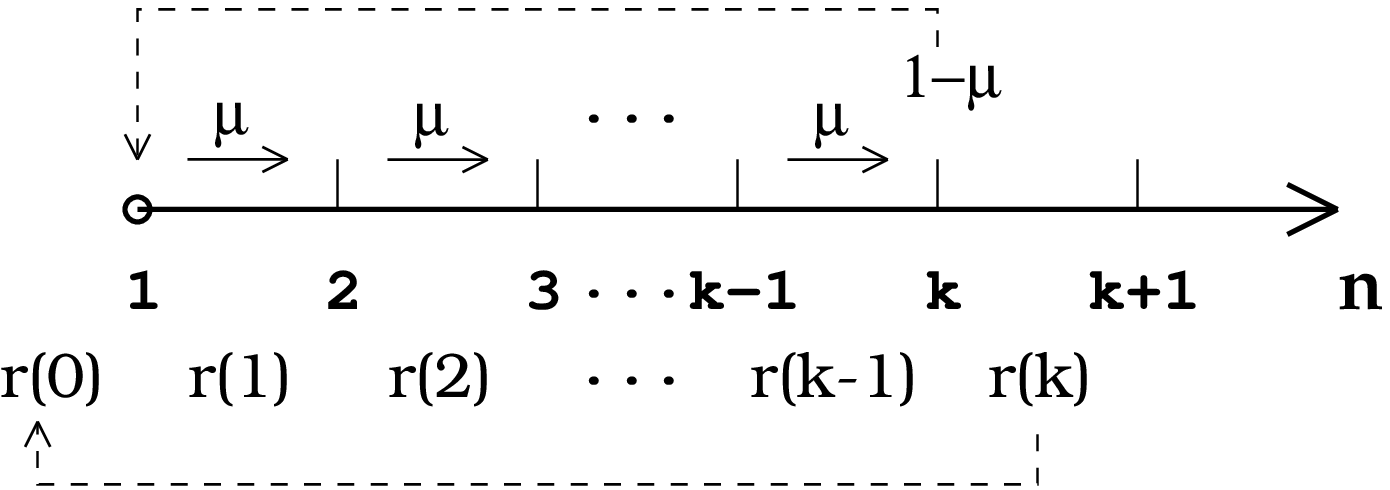}
\vskip 5mm
\caption{Graphical representation of the walk generation. A step is made
with probability $\mu$, a new walk initiated with probability $1-\mu$.
The random number $r(k)$ is used to decide whether the step is made
from $k$ to $k+1$.}
\label{walk}
\end{figure}

We may interpret the cluster formation in the simplified
model as a directed one-dimensional random walk.
A walk starts on one site (see Fig.~\ref{walk}).
At discrete times $n$ the walker steps to the neighbouring site in a
fixed direction (to the right in Fig.~\ref{walk}) with probability $\mu$.
Thus the walker cannot recur to a site once visited.
If the step is not made, a new walk is initialized.
The decision, whether to start a new walk or
not, does (ideally) not depend on previous decisions.
The probability to start a new walk at a certain time equals
$\nu = 1 - \mu$, independent of the length of the walk.
The probability $P(n)$ that this process generates a walk covering
precisely $n$ sites, satisfies
\begin{equation}
P(n)=\mu^{n-1}(1-\mu)
\label{eq_prob}
\end{equation}
and the expectation value of the number of sites visited is
\begin{equation}
\langle n \rangle=\frac{1}{1-\mu}
\end{equation}
One can formulate this random walk process in terms of
the following two algorithms.
We distinguish two possible ways to generate the decision
whether to step to the next site or to initialize a new walk .
The algorithms differ by the underlined statements.

\begin{minipage}{.5\textwidth}
\begin{tabbing}
\quad \=\quad \=\quad         \kill
Algorithm P \\
\\
Initialize walk statistics $\{n_L\}$\\
\\
generate a random number $r$\\
\> (with $0\leq r <1$)\\
\\
{\bf for} walk := 1 {\bf to} total number {\bf do} \\
\>  \{ walk length $L$:= 1\\
\> {\bf while} { $\underline{ \mu > r}$} {\bf do} \\
\> \> \{ $L:=L +1$\\
\> \> generate a random number $r$ \} \\
\> {\bf enddo}\\
\> $n_L:=n_L+1$ \} \\
{\bf enddo}\\
\\
output $\{n_L\}$\\
\end{tabbing}
\end{minipage}
\quad\begin{minipage}{.5\textwidth}
\begin{tabbing}
\quad \=\quad \=\quad         \kill
{Algorithm N} \\
\\
Initialize walk statistics $\{n_L\}$\\
\\
generate a random number $r$\\
\> (with $0\leq r <1$)\\
\\
{\bf for} walk := 1 {\bf to} total number {\bf do} \\
\>  \{ walk length $L$:= 1\\
\> {\bf while} $\underline{1-\mu \leq r}$  {\bf do} \\
\> \> \{ $L:=L +1$\\
\> \> generate a random number $r$ \} \\
\> {\bf enddo}\\
\> $n_L:=n_L+1$ \} \\
{\bf enddo}\\
\\
output $\{n_L\}$\\
\end{tabbing}
\end{minipage}

Using perfect random numbers, independent and
uniformly distributed between 0 and 1,
both algorithms would yield consistent and unbiased averages.
However, in real simulations, the random numbers are correlated,
as a consequence of the deterministic production rule.
This causes the distribution of $n$-tuples of consecutive
pseudo-random numbers to be non-uniform in the unit hyper-cube especially
for larger dimensionality $n$.
A certain degree of non-uniformity is acceptable for the numbers used
in the directed random walk algorithm.
It is sufficient,
that $n$-tuples with components larger or smaller than the quantity
used for the decision ($\mu$ in Algorithm $P$ and $1-\mu$
in Algorithm $N$) are present in the right proportions.
But even this weaker requirement is not satisfied in practice.
This affects the expectation values of the simulation results.
These deviations are discussed in Sections \ref{sec_core} and \ref{sec_fib}.

\section{Random generators}
\label{sec_rng}

In a numerical Monte Carlo calculation as
described in Section \ref{sec_model} one has to decide between
stepping with probability $\mu$ and initialization of a new walk
with probability $\nu \equiv 1-\mu$.
One needs a random sequence of two decision outputs,
where one occurs with the step probability $\mu$
and the other with the initialization probability $1-\mu$.
To this purpose we use pseudo-random numbers $r$
with $0\leq r <1$ and compare them to the relevant probability.

We generate the pseudo-random numbers with a number of rules.
One is the Generalized Feedback Shift Register (GFSR) method \cite{LP,KS}.
A sequence of pseudo-random integers $r(i)$,
represented by their binary expansion (usually 32 bits),
is generated by the rule
$r(i)=r(i-p)\oplus r(i-q_1)\oplus \cdots \oplus r(i-q_j)$, where the
exclusive-or operation $\oplus$ is applied bit-wise and the feedback
positions (lags) are ordered according to $p>q_1>q_2>\cdots >q_j$.
If $j=1$, we denote the feedback lag by $q$.
The leading bits (as well as all other bits) separately form a
sequence, that is generated by a feedback shift register.
The maximum length equals $2^p -1$.
Production rules that generate maximum-length bit-sequences,
are characterized by primitive polynomials \cite{Gol}.
Tables of primitive polynomials are listed in \cite{rule}.
The most widely used random number generator based on the GFSR method is
R250 \cite{KS} with $p=250$, $q=103$ and $r(i)=r(i-250)\oplus r(i-103)$.

Where primitive trinomials are known,
one can obtain other primitive polynomials by decimation
procedures \cite{Gol}.
For example, using every third number from a sequence generated with a rule
derived from a trinomial is equivalent to using a rule specified
by a pentanomial, which adds two terms to the primitive trinomial.
For a rule based on primitive polynomials of degree $p$
only sequences up to $p$ successive bits or numbers are independent.
In order to prevent problems one should avoid
unwanted initializations. For instance, if the $k$-th
bit of each initialized integer is zero, the same will hold for the
subsequently generated pseudo-random integers.

Furthermore we shall consider the lagged Fibonacci method:
$r(i)=r(i-p)+r(i-q)$, where the addition is understood
modulo $2^l$, where $l$ is the number of bits in a computer word.
The maximum sequence length\cite{Dai,Marsfib} is $2^{l-1} ( 2^p - 1)$.

\section{Examples of the correlation effect: Shift Registers.}
\label{sec_core}

An important factor contributing to the bias for a given random-number
generator is associated with the fact that, whenever a new walk is
initialized, the preceding decision was {\em not} to visit another site.
The last random number $r_P(0)$ used in the previous walk
in Algorithm $P$ thus obeys $r_P(0)\geq \mu$
(or obeys $r_N(0)<1-\mu=\nu$ in Algorithm $N$).
If the new walk visits another site, the next random number
$r_P(1) < \mu$ (or $r_N(1) \geq \nu$).
If the computer words consist of $l$ bits, only probabilities
that are a multiple of $2^{-l}$ are faithfully represented.
We suppose that the first $p$ numbers of a generalized
shift-register sequence may be considered as independent
(as if produced by an ideal random number generator),
i.e. the decisions with respect to the first $p-1$ steps 
occur with probability $\mu$ independent of the history.
The probability of a walk with less than $p$ steps
is that given by Eq. (\ref{eq_prob}).

However, deviations will occur for step number $p$.
The $p$-th number in the sequence depends on the
bits in $r(0)$ and those in the integer $r(p-q)$ on the feedback position.
This causes the initialization probability at this
point to differ from the value $\nu$ it should have.
We denote the actual probability by $\nu^*$.
This probability will depend on the chosen random generator
(GFSR or lagged Fibonacci, etc.).
In this Section we will focus attention on shift-register
random-number generators. The case of the lagged Fibonacci recipe
is treated in the next Section.

First, we analyze some simple examples. If $\mu=\frac{1}{2}$
just one random bit $b$ is needed to generate the decision,
where both outcomes occur with equal probability.
The random bit $b$ denotes the leading bit of the pseudo-random
number $r$.
The completion of the preceding walk implies, except for the first walk,
that $b(0)=1$ when Algorithm $P$ is used.
In the case that a decision is to be made on step $p$, the completion
of the preceding steps means that $b(i)=0$ for $0<i<p$.
Thus, irrespective of the number $j+1$ of feedback positions
$p,q_1,q_2,...,q_j$, $b(p)=1$
so that the walk ends, and a new one is initialized.
The actual initialization probability is denoted $\nu^*(p)$ and
satisfies $\nu^*_P(p)=1$ instead of the desired $\nu = 1- \mu $.
The maximum walk length is $p$ and occurs with a probability
of twice that expected from Eq. (\ref{eq_prob}).

In the case that a decision has to be made on step $p$ of
Algorithm $N$, $b(0)=0$ and $b(i)=1$ for $0<i<p$.
If the number of additional feedback bits is odd,
which is the case for maximum-length sequences,
$b(p)=1$ and the walker always proceeds to site $p$.
The actual initialization probability thus satisfies $\nu^*_N(p)=0$.
However, one has $b(p+1) = 0$, so that $\nu^*_N(p+1)=1$.
Thus all walks with a length exceeding $p-1$ have a length $p+1$.
Length $p$ occurs with probability 0, length $p+1$
with probability $P^*(p+1)={\left(\frac{1}{2}\right)}^{p-1}$.

These results can be generalized for all $\mu = 2^{-m}$ and
positive integer $m$.
The walk lengths will be smaller than $p+1$ in Algorithm $P$,
as the leading $m$ bits of $r(p)$ will be the same as those of $r(0)$.
For Algorithm $N$ the leading $m$ bits of $r(p+1)$ have to be zero.
Lengths larger than $p+1$ do not occur.

If $\mu > \frac{1}{2}$, much longer walks may occur, if the
GFSR-generator is used.
As an example of this regime we take $\mu = \frac{5}{8}$.
One needs three bits to make a decision.
The possible values are represented by the integers
$\tilde{r} \equiv \lfloor 8 r\rfloor$
where the brackets denote the integer part.

In the case of Algorithm $P$ a decision on step $p$ requires
$\tilde{r}(0) \geq 5$ and $\tilde{r}(i) <5 $ for $0<i<p$.
We assume that all admissible numbers for $\tilde{r}(i)$ with $i<p$
occur with equal probability, which strictly holds only for the
first walk.
For sequences generated
with a production rule derived from a primitive trinomial,
$\tilde{r}(p)$ equals 4, 5, 6 and 7 with probability $\frac{1}{5}$,
1, 2 and 3 with probability $\frac{1}{15}$
and 0 with probability 0.
This non-uniform distribution of $\tilde{r}$
leads to a probability $\nu^*_P(p)=\frac{3}{5}$ for the walk
to end at step $p$. This is different from the desired probability
$\nu=\frac{3}{8}$.
The probability of walks of length $p$ thus equals
$P^*(p)={\left(\frac{5}{8}\right)}^{p-1}\frac{3}{5}$.

For $\tilde{r}(i)$ with
\begin{equation}
p<i<\left\{\begin{array}{lll}
p+q& \mathrm{if}&q<\frac{p}{2}\\
2p-q& \mathrm{if}&q>\frac{p}{2}\\
\end{array}
\right.
\label{ineq}
\end{equation}
the numbers 4, 5, 6 and 7 occur with probability $\frac{2}{25}$,
the numbers 1, 2 and 3 with probability $\frac{4}{25}$
and the number 0 with probability $\frac{1}{5}$.
Then $\nu^*_P(i) = \frac{6}{25}$.
Chains of length $p+1$ occur with probability
$P^*(p+1)={\left(\frac{5}{8}\right)}^{p-1}\frac{2}{5}\cdot\frac{6}{25}$
provided $q\neq 1,p-1$.
Similar non-uniformities of the distribution lead to $\nu^*_P(p+q)=\frac{2}{5}$
or $\nu^*_P(2p-q)=\frac{2}{5}$.

For a production rule derived from a primitive pentanomial with
lags $p>q_1>q_2>q_3$ the actual initialization probability
$\nu^*_P(p)=\frac{57}{125}$ and
$\nu^*_P(i)=\frac{204}{625}$ for $i$ larger than $p$
and less than the minimum of $p+q_3$, $p+q_2-q_3$, $p+q_1-q_2$ and $2p-q_1$.
The deviations are less than for trinomials.

Whenever a decision on step $p$ is made in the case of Algorithm $N$,
the pseudo-random numbers satisfy $\tilde{r}(0) \leq 2$
and $\tilde{r}(i)>2$ for $0<i<p$.
Assuming that $\tilde{r}(i)$ with $i<p$
occurs with equal probability for all possibilities consistent with a walk
length larger than $p-1$,
$\tilde{r}(p)$ equals
4, 5, 6 and 7 with probability $\frac{1}{5}$,
1, 2 and 3 with probability $\frac{1}{15}$
and 0 with probability 0 in the case of a primitive trinomial.
So $\nu^*_N(p)=\frac{2}{15}$.
The probability of walks of length $p$ thus equals
$P^*(p)={\left(\frac{5}{8}\right)}^{p-1}\frac{2}{15}$.
For $\tilde{r}(i)$ with $i$ obeying inequality (\ref{ineq}) the values 4,
5, 6 and 7 occur with probability $\frac{2}{25}$,
the values 1, 2 and 3 with probability $\frac{4}{25}$
and the value 0 with probability $\frac{1}{5}$.
Thus $\nu^*_N(i)=\frac{13}{25}$.
Chains of length $p+1$  occur with probability
$P^*(p+1)={\left(\frac{5}{8}\right)}^{p-1}\frac{13}{15}\cdot\frac{13}{25}$,
if $q\neq 1,p-1$.
In a similar way $\nu^*_N(p+q)=\frac{37}{65}$
or $\nu^*_N(2p-q)=\frac{37}{65}$.

In the examples a resonance in $\nu^*(i)$ versus $i$ is found,
when the value of $r(0)$ affects the value of $r(i)$.
The first one encountered is a reflection of the production
rule itself;
for trinomials a relation between lags 0, $q$ and $p$.
The next one is a four-point correlation resulting
from the interference of the production rule and a shifted version
of the production rule.
The correlation between lags 0, $q$ and $p$ and that between lags
$q$, $2q$ and $p+q$ leads to a correlation between lags 0,
$2q$, $p$ and $p+q$.
Similarly the correlation between lags 0, $q$ and $p$ and that
between $p-q$, $p$ and $2p-q$ leads to a correlation between
0, $p-q$, $q$ and $2p-q$.
The deviation of the initialization probability
has a different sign for Algorithm $P$ and Algorithm $N$,
if the new correlation was between an odd number of lags.
It has the same sign for the four-point correlation.

If a walk is longer than $p$ in the case of Algorithm $P$
for a general real step probability $\mu > \frac{1}{2}$,
then $r_P(0)\geq \mu$ and $r_P(i) < \mu$ for $0<i<p$.
As a consequence of these inequalities,
the production rule generates $r(p)$ with
a non-uniform probability distribution.
For a GFSR-rule derived from a trinomial $r(p)=r(0)\oplus r(p-q)$.
The exclusive-or operation with a fixed operand is a permutation
of the numbers smaller than 1.
If both $r(0)$ and $r(p-q)$ are larger than $\mu$ the result
is less than $\frac{1}{2}$.
This implies that all possibilities for $r(p)> \mu$ are realized
for the reduced set of numbers $r(p-q)<\mu$.
So the probability that the walk ends, when it has visited $p$ sites,
is not equal to $\nu=1-\mu$, but
$\nu^*_P(p)=\frac{\nu}{\mu}$.
By similar arguments for $i$ obeying (\ref{ineq})
$\nu^*_P(i)=(2\mu-1)\nu / {\mu^2}$.

For Algorithm $N$ the number $r_N(0) <\nu$
and $r_N(i)\geq \nu$.
If $\nu = 2^{-m}$ with $m$ a positive integer, then
$r_N(0)$ has $m$ leading bits equal to zero, so $r_N(p)$ has the
leading $m$ bits the same as $r_N(p-q)$ and
$\nu^*_N(p)=0$.
A similar result is found for all lengths for which
a three-point correlation involving $r_N(0)$ is induced
by the production rule.
In particular in the case of a rule derived
from a primitive trinomial this holds for all lengths $2^k p$ with $k$
a non-negative integer, because a decimation by 2 leads to a
sequence generated by the same rule\cite{Gol}.
For general $\nu$ a more complicated argument leads to
\begin{displaymath}
\nu^*_N(p)=\frac{2(2^{n_b-l+1}-\nu)(\nu-2^{n_b-l})}{\nu\mu}
\end{displaymath}
and for $i$ obeying (\ref{ineq})
\begin{displaymath}
\nu^*_N(i)=\frac{\nu}{\mu}-
\frac{2(2^{n_b-l+1}-\nu)(\nu-2^{n_b-l})}{\nu\mu}
\end{displaymath}
with
\begin{displaymath}
n_b=\lceil \log_2(\nu 2^l +1) \rceil-1.
\end{displaymath}
where $\lceil x \rceil$ is the smallest integer $\ge x$.

The deviation of the actual initialization probability $\nu^*$
from the ideal value $\nu$ causes the probability of a walk to
visit $n$ sites $P^*(n)$ to deviate from the value for uncorrelated
numbers (\ref{eq_prob}).
We define
\begin{equation}
\delta P(n)\equiv \frac{P^*(n)}{P(n)}-1,\;
P^*(n)=\prod_{i=1}^{n-1}(1-\nu^*(i))\; \nu^*(n).
\label{dev_prob}
\end{equation}
For $i<p$ we expect no deviation, so $\delta P = 0$.
Substitution in (\ref{dev_prob}) of the values found for $\nu^*$ leads to
$\delta P_P(p)=\frac{\nu}{\mu}$ for Algorithm $P$.
With $i$ obeying inequality (\ref{ineq})
\begin{displaymath}
\delta P_P(i)=\frac{(2\mu-1)^2}{\mu^4}
\left(\frac{3\mu^2-3\mu+1}{\mu^3}\right)^{i-p-1}-1
\end{displaymath}

\begin{figure}
\epsfxsize=\textwidth
\epsffile{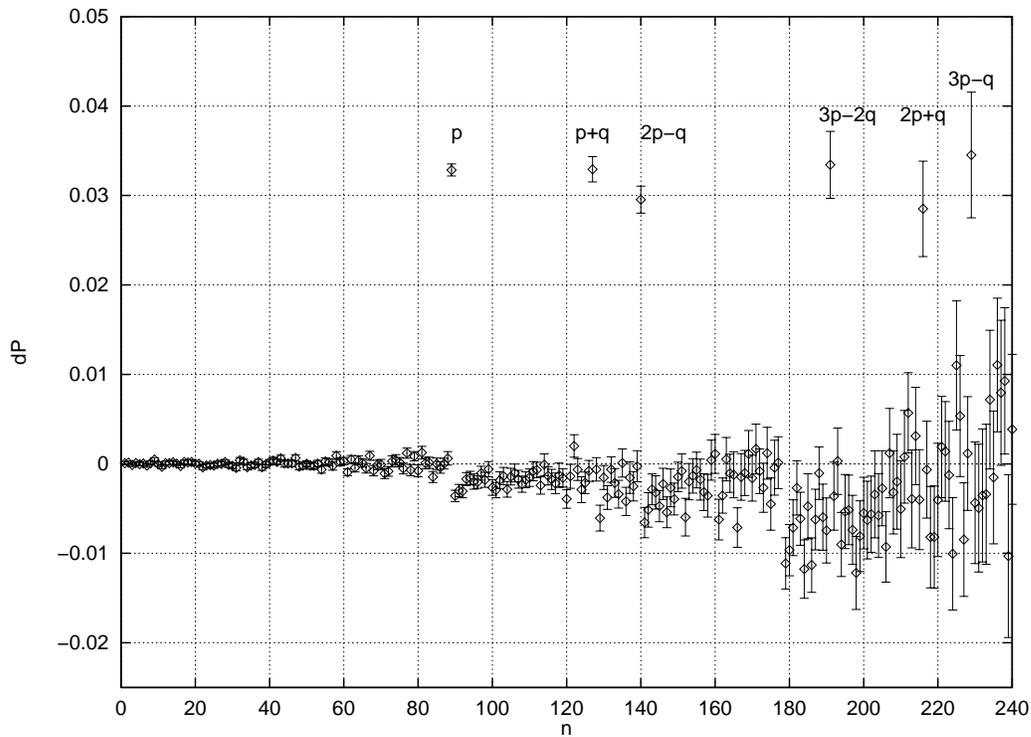}
\vskip 5mm
\caption{
Deviation $\delta P$ of the probability of a walk of length $n$
from the value for uncorrelated random numbers
versus walk length for a Monte Carlo calculation using Algorithm
$P$, $\mu=\frac{31}{32}$ and GFSR-rule (p,q)=(89,38).
Some of the resonances are labeled by linear combinations
of feedback positions.
The deviation at length $2p$ is equal to
$\delta P(2p)=0.109(3)$ and outside the scale of the figure.}
\label{fig_P}
\end{figure}

To give an impression of the deviations for lengths beyond
the first new low-order correlation
we use numerical calculations.
The results of computer simulations of $10^9$ walks are presented
in Fig.~\ref{fig_P} for the GFSR with $(p,q)=(89,38)$, Algorithm $P$
and $\mu=31/32$.
We mention two reasons to choose these particular values of $p$ and $q$.
First, simulations with the desired accuracy typically
need more than $10^{12}$ random numbers. Therefore, the length $p$ of the shift 
register  should be greater than $\log_2 10^{12} \approx 40$. 
Second, we do not want to choose $p$ much larger than necessary
because the observability of the bias decreases with $p$.

The error bars were computed using $100$ bins of $10^7$ walks each.
The predicted resonances are easily seen for the walk lengths
$p$, $p+q$, $2p-q$, $3p-2q$, etc., which are linear combinations
of the feedback positions in the production rule.
The result for walk length $2p$ is not shown in the Figure
because of its large deviation of $\delta P(2p)=0.109(3)$.
The value of the calculated deviation
$\delta P(p)=0.0328(7)$ is in good agreement with the value
$\frac{\nu}{\mu}\approx0.0323$.

\begin{figure}
\epsfxsize=\textwidth
\epsffile{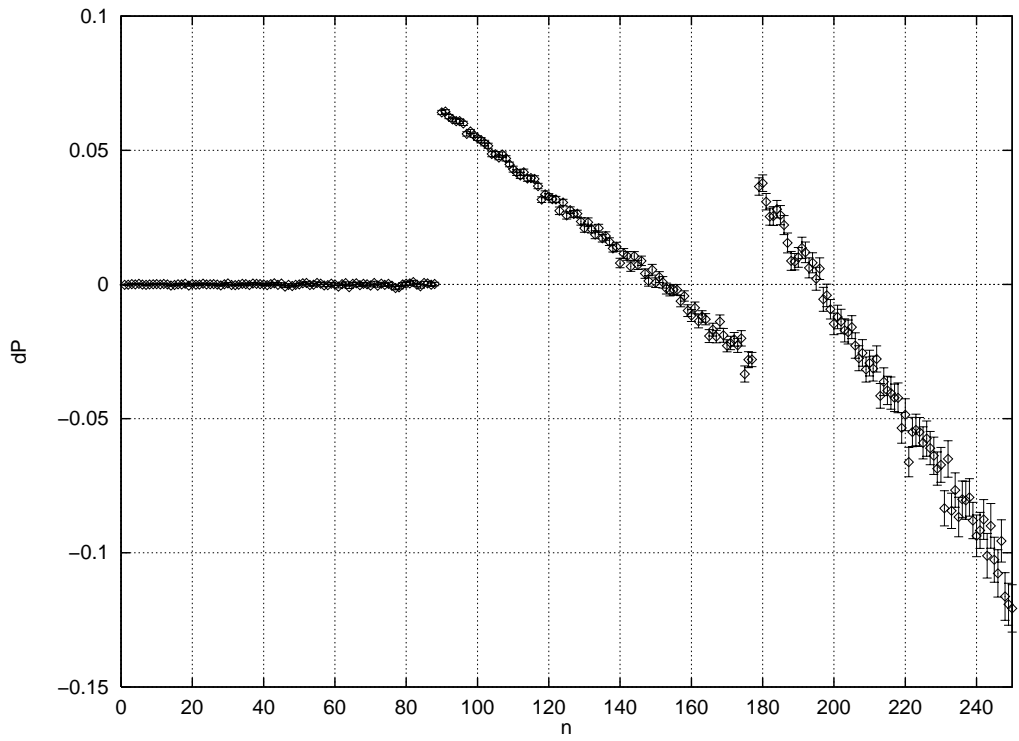}
\vskip 5mm
\caption{Deviation $\delta P$ of the probability of a walk
of length $n$ from the value for uncorrelated random numbers
versus walk length for a Monte Carlo calculation using
Algorithm $N$, $\mu=\frac{31}{32}$ and GFSR-rule (89,38).
The deviations are -1 for length $p$ and $2p$.}
\label{fig_N}
\end{figure}

The results of the calculations with the same parameters using
Algorithm $N$ are shown in Fig.~\ref{fig_N}. The deviations at lengths
equal to $2^kp$ ($k$=0,1,2...)
are equal to $-1$ and are not shown in the Figure.
The deviation $\delta P(p+1) = 0.0651(8)$ is in agreement with the
value $(1-\mu^2)/{\mu^2}\approx0.0656$.
A comparison of Fig.~\ref{fig_P} and
Fig.~\ref{fig_N} shows, that the main resonances have different signs
for the two algorithms.
This is in qualitative agreement with the results of cluster simulations
of the two-dimensional Ising model by Shchur and Bl\"ote \cite{SB} who
indeed found deviations in thermodynamical
quantities with opposite signs for comparison strategies $P$ and $N$.

\begin{figure}
\epsfxsize=\textwidth
\epsffile{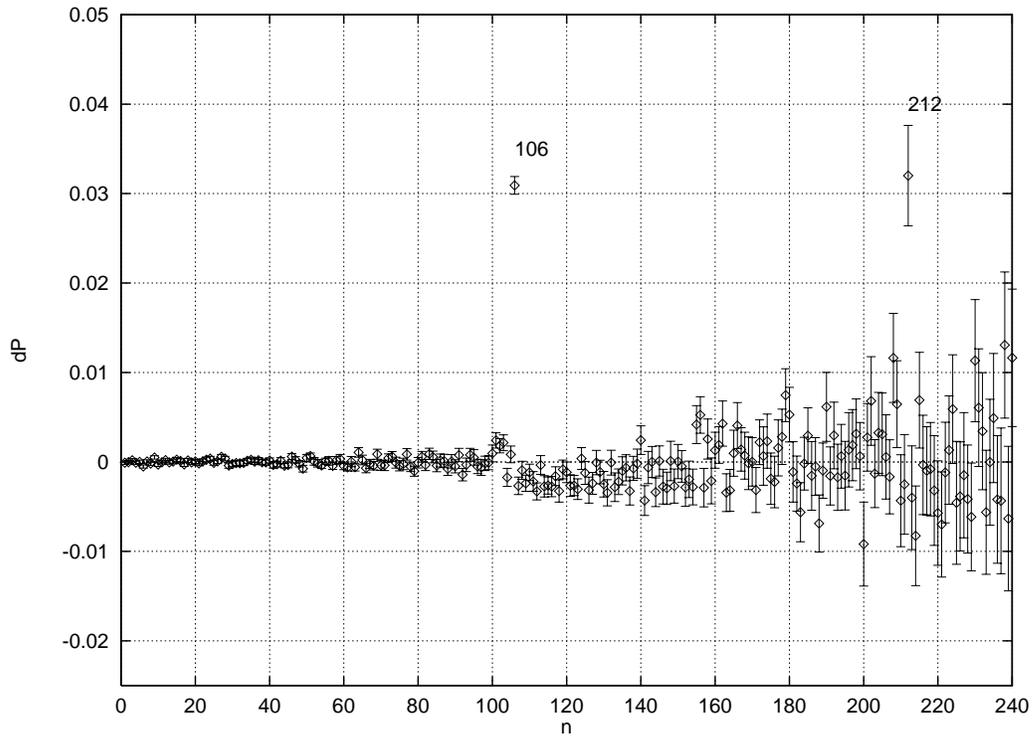}
\vskip 5mm
\caption{Deviation $\delta P$ of the probability of a walk
of length $n$ from the value for uncorrelated random numbers
versus walk length for a Monte Carlo calculation using
Algorithm $N$, $\mu=\frac{31}{32}$ and GFSR-rule (89,72,55,38).}
\label{fig_dec}
\end{figure}

In order to give an example of the deviations of random walk statistics
for a more complicated production rule we use the decimated sequence with
($p$,$q_1$,$q_2$,$q_3$)=(89,72,55,38), Algorithm $P$ and $\mu=\frac{31}{32}$
in a random walk simulation.
Because the feedback positions for this sequence are equally spaced,
the four numbers with lags 106, 38, 17 and 0 are correlated.
This four-point correlation may be more important in applications
than the five-point correlation implied by the production rule.
This is the case for the numerical results of the random walk statistics
(Fig.~\ref{fig_dec}).
The deviation at the shift-register length $\delta P(p)=1/{31^3}$
is small compared to the error bars.
The deviations for the two four-point correlations at walk lengths
106 and 212 are of a similar magnitude as the deviations for
four-point correlations for the trinomial (Fig.~\ref{fig_P}).
The number of four-point correlations is smaller than for the trinomial
and no three-point correlations are found.
Deviations tend to be weaker for higher order correlations.
This agrees with results of Wolff simulations of Ising models\cite{SB},
where the deviations were smaller for decimated sequences.

\section{Correlation effects due to the lagged Fibonacci recipe.}
\label{sec_fib}

For lagged Fibonacci generators the relation between the leading bits
is less simple.
The distribution of numbers is not completely uniform
for lagged Fibonacci sequences.
Deviations of uniformity have been analysed in \cite{Pinch}.
We assume that all sequences $r(0)\ldots r(p-1)$ are equally probable.
If the walk has a certain length the previous numbers $r(i)$ obey
the appropriate inequality.

We first consider $\mu=\frac{1}{2}$ again.
For Algorithm $P$, $r(0) \geq \frac{1}{2}$
and $r(i) < \frac{1}{2}$ for $0<i<p$, in the case of a decision
on step $p$.
These inequalities cause the production rule to generate $r(p)$ with a
non-uniform probability distribution.
The distribution can be calculated as the convolution of the distributions
for the feedback lags.
For numbers of infinite precision it has
the symmetry property $P(r)=P(1-r)$
in the case of a rule derived from a primitive trinomial.
The finite word length $l$ leads to corrections of order $2^{-l}$.
We neglect these corrections because they are small compared to the
statistical errors in our simulations.

The actual initialization probability conserves its ideal value
$\nu^*_P(p) = \frac{1}{2}$ by the symmetry property.
The same symmetry holds for the distribution of $r(i)$
with $i$ obeying inequality (\ref{ineq}).
Thus $\nu^*_P(i) = \frac{1}{2}$.
For the next random number the probability distribution
does no longer obey $P(r)=P(1-r)$.
The actual initialization
probability $\nu^*_P(p+q)=\frac{1}{3}$
or $\nu_P^*(2p-q)=\frac{1}{3}$.
The same results are obtained for algorithm $N$.

As in the case of the GFSR rule it is possible to calculate the
values of the actual initialization probability for arbitrary $\mu$.
For a lagged Fibonacci rule derived from a trinomial the actual
initialization probability becomes
$\nu^*_P(p)=\frac{\nu}{2 \mu}$.
For $i$ obeying inequality (\ref{ineq}) the actual probability
$\nu^*_P(i)=(3\mu-1)\nu / {2\mu^2}$.
For strategy $N$ we get the same actual probabilities.
The expressions for $\nu^*$ grow progressively more
complicated for larger walk lengths.
We therefore refrain from showing them.

\begin{figure}
\epsfxsize=\textwidth
\epsffile{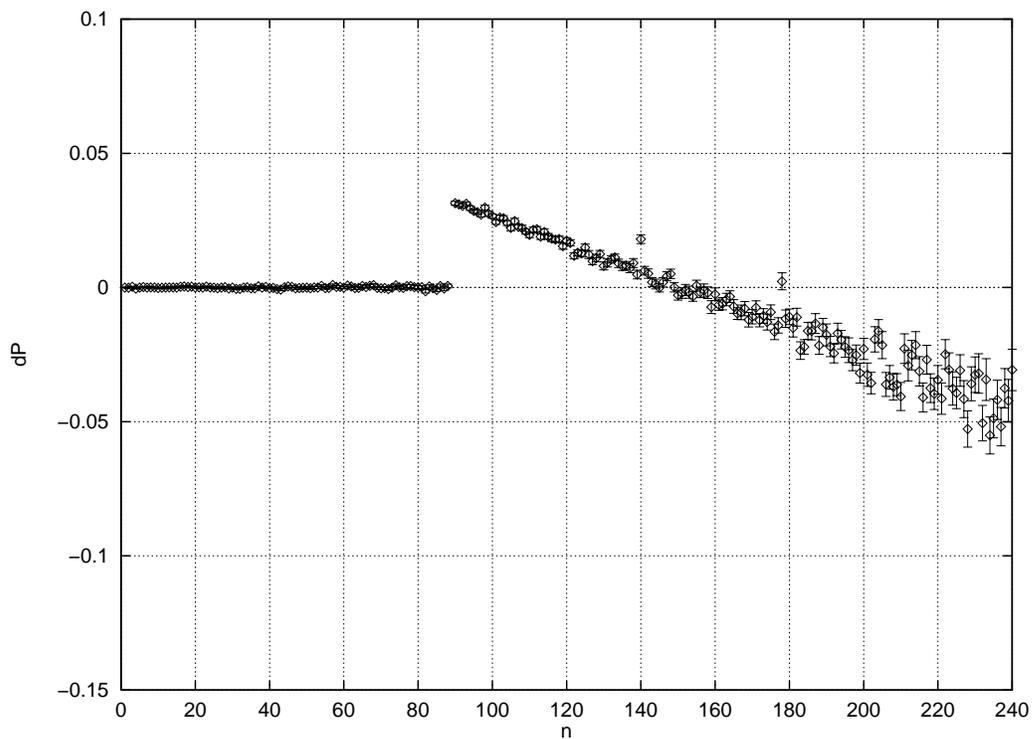}
\vskip 5mm
\caption{Deviation $\delta P$ of the probability of a walk of length $n$
from the value for uncorrelated random numbers
versus walk length for a Monte Carlo calculation using
Algorithm $P$, $\mu=\frac{31}{32}$ and
lagged Fibonacci rule (89,38).
The deviation at length $p=89$ equals -0.4841(4) and is not shown in the
figure.}
\label{fig_F}
\end{figure}

Numerically calculated deviations of the resulting probability
$P^*(n)$ of a walk of length $n$ are shown
in Fig.~\ref{fig_F} for the lagged Fibonacci rule with feedback positions
$(p,q)=(89,38)$. Error bars were computed using the same parameters
as in the previous Section.
Because of the number of bits $l=30$ used in the random numbers
the finite-word-length corrections are quite small.
As those differences are small compared to the statistical errors,
the results for Algorithms $P$ and $N$ are equal within error bars
for all walk lengths, unlike in the case of shift registers.
The value of the deviation at walk length
$p$ (not shown in the Figure)
is equal to $\delta P(p)=-0.4841(4)$ in good agreement with the
value of $\frac{1}{2\mu}-1\approx -0.4839$.
The deviations have the same order of magnitude for both recipes of
random number generation we considered in this paper.
This does not support the claim that the lagged Fibonacci method
performs better than the GFSR-method\cite{Marsfib}.

\section{Conclusions}
\label{sec_dis}

Correlation between random numbers can influence the
results of a Monte Carlo calculation of a simple random walk model.
No deviations occur in the distribution of walk lengths
shorter than the magnitude of the largest feedback position.
The walk length statistics is affected for larger lengths.
In some cases the difference in results using comparison
strategy $N$ and $P$ gives an indication of the magnitude
of the bias.
The magnitude of the deviations tends to be larger for generators
derived from primitive trinomials than for primitive pentanomials.
Similar effects occur in the cluster-size distribution
in a cluster simulation of the Ising model\cite{LNS}.

The magnitude of the deviations $\nu - \nu^*$ depends only on
the comparison strategy, and on the value $\mu $.
It does not depend on the particular values of feedback positions.
In the case of other feedback positions, for example R250 the
Kirkpatrick-Stoll Random Number Generator, 
one should accordingly rescale the horizontal axes of Figures 2-4 by
$250/89$.
However, continuation of the walks to larger $n$ leads to increased
scatter so that the effects become less prominent.

These results are relevant for the Wolff cluster simulation of spin
models in more than one dimension. In such simulations, each spin in
the cluster may have more than one neighbour that has to be included
in the cluster. If so, the addresses of these neighbouring spins are
temporarily stored, e.g. in a memory called `stack'( for a hardware
implementation see \cite{cspc}). For each entry in the
stack, it has to be checked whether further neighbours have to be added
to the cluster, and thus whether further additions to the stack memory
have to be made. Spin addresses that have thus been processed,
are removed from the stack.
The number of addresses in the stack is a fluctuating
variable, and the cluster is completed when the stack is empty.
Typically the stack memory contains more than one spin address, in which
case one random number is not sufficient to end the cluster formation
process. However, it is obvious that the completion of a cluster is
strongly correlated with the values of the preceding pseudo-random
numbers.

The bias in the random numbers preceding the construction of a new
cluster will lead to a significant correlation between $n-1$
pseudo-random numbers in the case of an $n$-point production rule,
although the correlation is weaker than the correlation between
$n$ numbers imposed by the production rule.
For a 3-point rule this 2-point correlation combined with the bias
in the random numbers allowing a further growth of the cluster causes
a bias for the $p$-th pseudo-random number used in the construction of
a cluster.
It is known\cite{SB} that the bias in the simulation results
(for three-point production rules)
becomes largest, when $p$ is equal to the average cluster size.

For higher $n$ the bias-producing mechanism is less simple and it seems
plausible that the bias in the $p$-th pseudo-random number will be weaker.
Thus one expects that the deviation in the statistics of cluster sizes
will be stronger,
when less pseudo-random numbers are involved in the production rule.
Indeed for the case of shift-registers with a $n$-point production rule,
the bias in the thermodynamics in a Wolff simulation increases
strongly with decreasing $n$ \cite{SB}, where rules with higher values
of $n$ can be generated either by decimation or by combining the
numbers generated by two or more rules with an exclusive-or.

\begin{ack}
We acknowledge productive discussions with A. Compagner,
W. Selke, D. Stauffer and A.L. Talapov.
L.N.S. is grateful to the Computational Physics Group
at TU-Delft, where this work was initiated, for their kind
hospitality. This work is partially supported by grants
RFBR 93-02-2018, NWO 07-13-210, INTAS-93-211.
\end{ack}

\newpage

\end{document}